\documentclass[aps,prd,reprint,nofootinbib,superscriptaddress]{revtex4-2}
 \usepackage{bm}
 \usepackage{indentfirst}
 \usepackage{amsmath}
 \usepackage{graphicx}
 \usepackage{float}
 \usepackage{amssymb}
 \usepackage{subfigure}
 \usepackage{amssymb}
 \usepackage{hyperref}
 \hypersetup{
 	colorlinks=true,
 	linkcolor=red,
 	citecolor=blue,
 }

\begin{document}

 \title{Primordial black holes and secondary gravitational waves from the Higgs field}

 \author{Zhu Yi)}
 	\email{yz@bnu.edu.cn}
 	\affiliation{Department of Astronomy, Beijing Normal University,
 	Beijing 100875, China}
 	\author{Yungui Gong}
 	\email{Corresponding author. yggong@hust.edu.cn}
 	\affiliation{School of Physics, Huazhong University of Science and Technology, Wuhan, Hubei 430074, China}
 \author{Bin Wang}
\email{wang\_b@sjtu.edu.cn}
\affiliation{School of Aeronautics and Astronautics, Shanghai Jiao Tong University, Shanghai 200240, China}
\affiliation{Center for Gravitation and Cosmology, Yangzhou University, Yangzhou 225009, China}
  	\author{Zong-hong Zhu}
 	\email{zhuzh@bnu.edu.cn}
 	\affiliation{Department of Astronomy, Beijing Normal University,
 	Beijing 100875, China}
 	
\preprint{2007.09957}

\begin{abstract}
We devise a novel mechanism and for the first time
demonstrate that the Higgs  model in particle physics can drive the inflation to satisfy the cosmic microwave background observations and simultaneously enhance the curvature perturbations at small scales to explain the abundance of dark matter in our universe in the form of primordial black holes.
The production of primordial black holes is  accompanied
by the secondary gravitational waves induced by the first order Higgs fluctuations which is expected observable by space-based
gravitational wave detectors. We propose possible cosmological probes of Higgs
field in the future observations for primordial black holes dark matter or stochastic gravitational waves.

\end{abstract}
 	
\maketitle


\section{Introduction}	

The detections of gravitational wave (GW) by the
Laser Interferometer Gravitational Wave Observatory
(LIGO) Scientific Collaboration and the Virgo
Collaboration \cite{TheLIGOScientific:2016agk,Abbott:2016blz,Abbott:2016nmj,Abbott:2017vtc,
Abbott:2017oio,TheLIGOScientific:2017qsa,Abbott:2017gyy,LIGOScientific:2018mvr,
Abbott:2020uma,LIGOScientific:2020stg,Abbott:2020khf,Abbott:2020tfl,Abbott:2020niy}
started a new era of multimessenger astronomy. It was argued that GW observations can disclose the property of primordial black holes (PBHs) \cite{Bird:2016dcv,Sasaki:2016jop} which could explain dark matter (DM)  \cite{Ivanov:1994pa,Frampton:2010sw,Belotsky:2014kca,Khlopov:2004sc,Clesse:2015wea,
Carr:2016drx,Inomata:2017okj,Garcia-Bellido:2017fdg,Kovetz:2017rvv,Carr:2020xqk}.
PBHs can be formed through gravitational collapse of highly overdense inhomogeneities with  density contrast
exceeding the threshold value at the horizon reentry
in the radiation era \cite{Carr:1974nx,Hawking:1971ei}.
Such large density contrast can arise from the primordial curvature perturbations
during inflation \cite{Gong:2017qlj,Martin:2012pe,Motohashi:2014ppa,Garcia-Bellido:2017mdw,Germani:2017bcs,Motohashi:2017kbs,Ezquiaga:2017fvi,Bezrukov:2017dyv,Espinosa:2017sgp,Ballesteros:2018wlw,Sasaki:2018dmp,Kamenshchik:2018sig,Gao:2018pvq,Dalianis:2018frf,Dalianis:2019vit,Passaglia:2018ixg,Passaglia:2019ueo,Fu:2019ttf,Fu:2019vqc,Xu:2019bdp,Braglia:2020eai,Gundhi:2020zvb,Fumagalli:2020adf,Fumagalli:2020nvq,Lin:2020goi,Yi:2020cut,Gao:2020tsa,Gao:2021vxb}.
However, the constrained amplitude of the scalar power spectrum from Planck 2018 measurements
of the cosmic microwave background (CMB) anisotropy is $A_s=2.1\times 10^{-9}$
at the pivot scale $k_*=0.05\ \text{Mpc}^{-1}$ \cite{Akrami:2018odb}.
To produce abundant PBH DM, the amplitude of the
scalar power spectrum is required to reach
the order $A_s\sim \mathcal{O}(0.01)$ \cite{Lu:2019sti,Sato-Polito:2019hws},
so we need  mechanisms to enhance the
curvature power spectrum at small scales by seven orders of magnitude of the CMB observed value. Such enhancement also induces secondary GWs  after the horizon reentry
\cite{Matarrese:1997ay,Mollerach:2003nq,Ananda:2006af,Baumann:2007zm,Garcia-Bellido:2017aan,Saito:2008jc,Saito:2009jt,Bugaev:2009zh,Bugaev:2010bb,Alabidi:2012ex,Orlofsky:2016vbd,Nakama:2016gzw,Inomata:2016rbd,Cheng:2018yyr,Cai:2018dig,Bartolo:2018rku,Bartolo:2018evs,Kohri:2018awv,Espinosa:2018eve,Cai:2019amo,Cai:2019elf,Cai:2019bmk,Cai:2020fnq,Domenech:2019quo,Domenech:2020kqm}.
Therefore, the observations of PBH DM and scalar induced secondary GWs (SIGWs) provide novel probes of  physics in the early universe.

Assuming the Higgs boson as inflaton with the potential $\lambda\phi^4/4$,
we find too small
 scalar spectral tilt $n_s$  and too big tensor-to-scalar ratio $r$ to be allowed by CMB observations.
To reduce $r$,
Higgs inflation introduces the nonminimal coupling $\xi\phi^2 R$ between Higgs field and gravity \cite{Kaiser:1994vs,Bezrukov:2007ep},
New Higgs inflation introduces the nonminimally derivative coupling $G^{\mu\nu}\partial_\mu\phi\partial_\nu\phi/M^2$
between the kinetic term of the Higgs field and Einstein tensor $G^{\mu\nu}$
\cite{Germani:2010gm,Germani:2014hqa, Yang:2015pga,Fumagalli:2017cdo,Fumagalli:2020ody}
and Gauss-Bonnet inflation introduces a special relation between the inflationary potential and the coupling between the inflaton and the Gauss-Bonnet term \cite{Yi:2018gse} to reconcile the observables $n_s$ and $r$ to
satisfy CMB observations.
The nonminimal coupling may have signatures of the
so called Higgs shifts near strong gravity sources \cite{Onofrio:2010zz,Onofrio:2014txa,Wegner:2015aea}.
However, these nonminimal couplings cannot provide large enough curvature perturbations at small scales.
Adopting the observed values of Higgs boson and top quark masses,
the coupling $\lambda$ in the Higgs potential is allowed to become negative
from the running of the Higgs self-coupling via the renormalization group equations.
In critical Higgs inflation \cite{Hamada:2014iga,Bezrukov:2014bra}, near the critical point
$\lambda=\beta_\lambda=0$, the curvature power spectrum can be enhanced around
the inflection point in the Higgs potential \cite{Ezquiaga:2017fvi},
but such an enhancement is only five orders of magnitude of the CMB measurement and is unable to produce significant abundance of PHBs \cite{Bezrukov:2017dyv}.
In the spectator Higgs model,
when the Higgs field stays in the unstable phase of the Higgs potential during inflation,
the quantum fluctuations of the Higgs field can produce abundant PBH DM \cite{Espinosa:2017sgp}.
In this mechanism, the Higgs field is not responsible for inflation.
However it was found that Higgs fluctuations on CMB scales
are larger than those on PBH scales \cite{Passaglia:2019ueo}.

In a single field inflation, it was claimed difficult to enhance the
amplitude of the power spectrum to the order $\mathcal{O}(0.01)$
while keeping the total $e$-folding number  $N\simeq 50-60$ \cite{Sasaki:2018dmp,Passaglia:2018ixg}.
Generalizing the coupling $1/M^2$ to
a special function $g(\phi)=d/\sqrt{1+(\phi-\phi_{r})^2/c^2}$ in the
nonminimally derivative coupling, an enhancement of the
CMB power spectrum up to seven orders of magnitude
at small scales was achieved, but the  price to pay is to restrict
the potential to be in the specific form $\phi^{2/5}$ \cite{Fu:2019ttf}.
Similar to k inflation \cite{ArmendarizPicon:1999rj,Garriga:1999vw} and G inflation \cite{Kobayashi:2010cm},
the noncanonical kinetic term was proposed to increase the curvature perturbations at small scales and achieve abundant production of PBH DM and SIGWs in k/G inflation \cite{Lin:2020goi,Yi:2020cut,Gao:2020tsa,Gao:2021vxb}.
In this mechanism the noncanonical kinetic term succeeds  enhancing  the perturbation power spectrum
at small scales while keeping such effect negligible at large scales,
but the potential form is also restricted.
It is fair to conclude that so far there is no available Higgs mechanism
that can successfully satisfy observational requirements of inflation at large scales and simultaneously enhance the power spectrum at small scales.
The reasonable question we intend to ask is whether the standard Higgs
field model can drive inflation and produce abundant PBH DM without introducing
other fields beyond the standard model.
In this paper, we devise a novel mechanism in the framework of a single field inflation with Higgs potential to enhance the primordial curvature perturbation at small scales while keeping it negligible at large scales. We show that this model is consistent with Planck 2018 data and produces a significant abundance of PBH DM and SIGWs to be detected by the future space-based GW detectors such as LISA \cite{Danzmann:1997hm,Audley:2017drz},
TianQin \cite{Luo:2015ght}, and Taiji \cite{Hu:2017mde}.
In our mechanism, the Higgs field not only drives inflation
but also is responsible for the PBH DM content of our universe.
It is interesting to note that our mechanism does not only work for the Higgs field,
it is a general single field inflationary model to explain the abundance of PBH DM and can be generalized to other inflationary models, for example the T-model.

\section{The enhancement mechanism}

For a slow-roll inflation with the noncanonical kinetic term $[1+G(\phi)]\dot\phi^2/2$, the power spectrum of the primordial curvature perturbation is
\begin{equation}
P_\zeta=\frac{H^4}{4\pi^2\dot{\phi}^2(1+G)}\approx \frac{V^3}{12\pi^2V_{\phi}^2}(1+G),
\end{equation}
where we choose $M_\text{pl}=1/\sqrt{8\pi G}=1$, $V_\phi=dV/d\phi$ and
the noncanonical kinetic term may arise from scalar-tensor theory of gravity, G inflation \cite{Kobayashi:2010cm} or k inflation \cite{ArmendarizPicon:1999rj,Garriga:1999vw}.
If the function $G(\phi)$ has a peak, then the power spectrum can be enhanced.
Motivated by the $\omega(\phi)=1/\phi$ coupling in Brans-Dicke theory \cite{Brans:1961sx}, the function
\begin{equation}
\label{Ga}
G(\phi)=G_a(\phi)=\frac{h}{1+|\phi-\phi_p|/w},
\end{equation}
is used to enhance the power spectrum so as to produce abundant PBH DM and observable SIGWs \cite{Lin:2020goi},
where $h \sim \mathcal{O}(10^{10})$ gives the amplitude of the peak,
$w\sim \mathcal{O}(10^{-10})$ controls the width of the peak
and the number of $e$-folds before the end of inflation
at the horizon exit for the pivotal scale,
$\phi_p$ determines the position of the peak which is related
with the peak mass of PBH and the peak frequency of SIGWs.
Away from the peak, $|\phi-\phi_p|/w\gg 1$, the function $G_a(\phi)$
becomes negligible and the usual slow-roll inflation resumes.
At the horizon exit, the number of $e$-folds remaining in inflation is
\begin{equation}\label{efold0}
 N=\int_{\phi_e}^{\phi_*} (1+G)\frac{V}{V_{\phi}}d \phi,
\end{equation}
where $\phi_*$ is the field value at the horizon exit and $\phi_e$
is the field value at the end of inflation.
The peak in the noncanonical kinetic term $G(\phi)$ contributes up to $\sim 20$ $e$-folds, which effectively moves $\phi_*$ closer to $\phi_e$
in order to keep the total number of $e$-folds around 60. The effective $e$-folds contributed by the standard slow-roll inflation then reduces to around 40, so that $n_s$ and $r$  in this mechanism
become incompatible with CMB observations, if we choose allowed inflationary potentials in standard canonical models, but this is the price to pay for the enhancement of the power spectrum at small scales due to the noncanonical coupling $G(\phi)$.
In particular, this mechanism does not work for the Higgs field.

Taking the advantage of the enhancement mechanism in k/G inflation,
we invent a new coupling function $f(\phi)$ which has the chameleon effect to keep the enhancement of
the curvature perturbations at small scales,
while at large scales it can adjust the predictions of $n_s$ and $r$  to meet Planck 2018 data.
The noncanonical term, which might come from some kinds of scalar-tensor theory of gravity, becomes
\begin{equation}\label{coupling}
G=G_a+f(\phi).
\end{equation}
At the end of the inflation, the scalar field rolls down to its minimum, the
noncanonical term becomes negligible.
In our new mechanism, the function $G_a(\phi)$
is general but not restricted to the form in Eq. \eqref{Ga}, see Refs. \cite{Yi:2020cut,Gao:2020tsa,Gao:2021vxb} for other functions of $G_a(\phi)$.
Introducing the function $f(\phi)$ we can modify the shape of the potential.
Away from the peak $\phi_p$, the effect of $G_a(\phi)$ is negligible
and the function $f(\phi)$ dominates.
We can change the noncanonical field $\phi$ to the canonical field
$\Phi$ by the transformation $d\Phi=\sqrt{f(\phi)}d\phi$.
In terms of the canonical field, the potential is $U(\Phi)=V[\phi(\Phi)]$.
To show how the mechanism works, without loss of generality, we take the potential $U(\Phi)$ in a power law form $U(\Phi)=U_0\Phi^n$, so
\begin{gather}
\label{cns}
n_s=1-\frac{n+2}{2N},\\
\label{cr}
r=\frac{4n}{N}.
\end{gather}
Without the enhancement at small scales, $N\sim 60$, it is easy to see that
no chaotic inflation is consistent with observational constraints.
However with the enhancement,
the effective number of $e$-folds $N$ for the canonical field
is around 40.
Taking $n=1/3$, we get $n_s=0.971$ and $r=0.033$.
If we take $n=2/3$, we get $n_s=0.967$ and $r=0.067$. Therefore,
depending on the function $G_a(\phi)$ and the model parameters,
it is possible that the predictions of these models are consistent with CMB
constraints $n_s=0.9649\pm 0.0042$ (68\% CL) and $r_{0.05}<0.06$ (95\% CL) \cite{Ade:2018gkx}.
Given the power law form for $U(\Phi)$ and $V(\phi)$, we can get
the function $f(\phi)$,
\begin{gather}
\label{gfuncn}
f(\phi)= \frac{1}{n^2}\left(\frac{1}{U_0}\right)^{2/n}V^{\frac{2}{n}-2}V_\phi^2.
\end{gather}
From the above argument, we see that our mechanism does not restrict the form of
the potential and it works for T-model and natural inflation as explored in Refs. \cite{Yi:2020cut,Gao:2020tsa}.

\begin{figure}
	\centering
	\includegraphics[width=0.95\columnwidth]{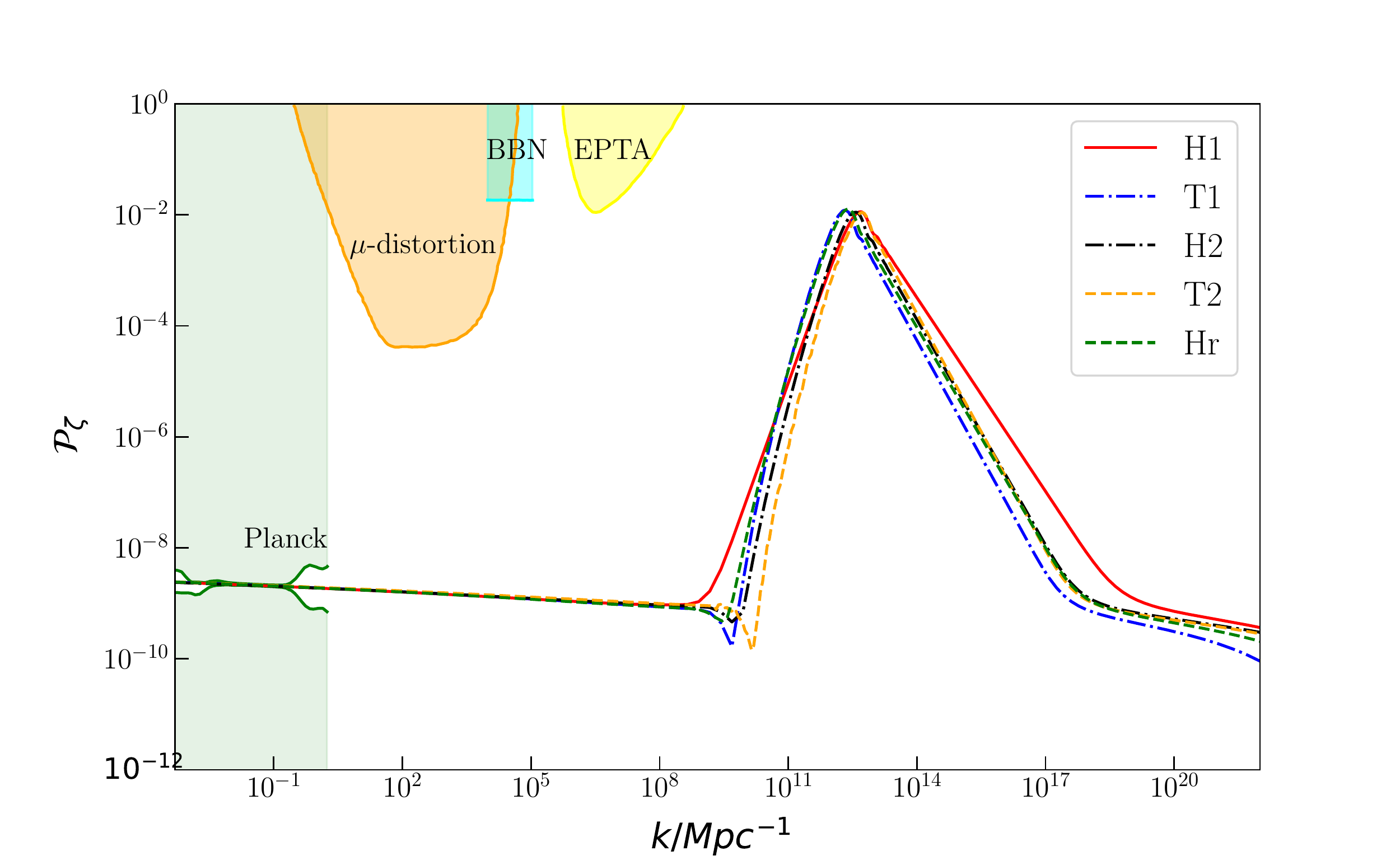}
	\caption{The results for the power spectrum of primordial curvature perturbations
along with the observational constraints. H represents the Higgs potential and T represents the T-model, 1 represents the case $n=1/3$ and 2 represents the case $n=2/3$. H1 means the model with Higgs potential and the power law potential $U(\Phi)=U_0\Phi^n$ with $n=1/3$.
	The light green shaded region is excluded by the CMB observations \cite{Akrami:2018odb}. The yellow, cyan and orange regions show the
    constraints from the PTA observations \cite{Inomata:2018epa},
    the effect on the ratio between neutron and proton
    during the big bang nucleosynthesis (BBN) \cite{Inomata:2016uip}
    and $\mu$-distortion of CMB \cite{Fixsen:1996nj}, respectively.
    }
	\label{p1}
\end{figure}

For the Higgs potential $V=\lambda\phi^4/4$,
we first take the model $U(\Phi)=U_0\Phi^n$ with $n=1/3$
as an example and label it as H1.
In this case, the function $f(\phi)=f_0\phi^{22}$ with $f_0=9(\lambda/U_0)^6/256$ and we choose $f_0=1$.
In effective field theory, lower dimensional terms $\sum_{4}^{21} C_n \phi^n$ with $C_n<O(1)$ also appear in $f(\phi)$.
It was found that the lower dimensional terms have little effect on this enhancement mechanism \cite{Yi:2020cut}.
In the low energy regime after inflation $\phi\ll M_{\text{pl}}=1$, the Higgs field runs away from
the peak and the function $f(\phi)$ becomes negligible leading to
the recovery of the canonical field.
Choosing the parameters $h$, $w$, $\phi_*$, $\phi_p$ and $\lambda$
as shown in Table \ref{t1},
and solving the equations for the background and the perturbations numerically,
we get $n_s=0.968$, $r=0.0383$ and $N=62.3$.
The chosen parameter sets and the results
are shown in Tables \ref{t1} and \ref{t2}. The power spectrum of the primordial curvature perturbations is shown in Fig. \ref{p1}.

\begin{table*}[htp]
	\begin{tabular}{llllllllll}
		\hline
		\hline
		Model  \quad  & $h$& $w$ &$\phi_p$&$\phi_*$&$\lambda /V_0$ &$N$ & $n_s$& $r$ &$k_\text{peak}/\text{Mpc}^{-1}$  \\
		\hline
		H1 & $1.05\times 10^{10}$ & $2.04\times 10^{-10}$ & $1.344$ & $1.40$ & $1.24\times 10^{-9}$  & 62.3 & 0.9680 & 0.0383 & $4.86\times 10^{12}$\\

		H2 & $7.13\times 10^9$& $1.94\times 10^{-10}$ &$1.750$&$1.88$&$6.40\times 10^{-10}$ & 64.2 & 0.9697 & 0.0641 & $3.67\times 10^{12}$ \\
		
		Hr &$2.14\times 10^{12}$& $1.13\times 10^{-11}$ & 0.1505 & 0.157 & & 60.2 & 0.9676 & 0.0540 & $2.54\times 10^{12}$ \\\hline
		
		T1 &$4.72\times 10^9$& $8.89\times 10^{-11}$ &$0.451$&$0.81$&$1.67\times 10^{-9}$ & 55.6& $0.9687$ & $0.0370$ &$2.17\times 10^{12}$ \\

		T2 &$8.90\times 10^9$& $4.75\times 10^{-11}$ &$0.835$&$1.35$&$2.95\times 10^{-9}$&$63.4$&$0.9704$& $0.0598$ &$5.24\times 10^{12}$ \\
		\hline
		\hline
	\end{tabular}
	\caption{The chosen parameter sets and the results. H represents the Higgs potential and T represents the T-model, 1 represents the case $n=1/3$ and 2 represents the case $n=2/3$. H1 means the model with Higgs potential and the power law potential $U(\Phi)=U_0\Phi^n$ with $n=1/3$. Hr represents the Higgs model with the running of the coupling constant $\lambda$.}
\label{t1}
\end{table*}

\begin{table}[htp]
	\begin{tabular}{lllll}
		\hline
		\hline
		Model \quad   &$P_{\zeta(\text{peak})}$&$M_\text{peak}/M_\odot$&$Y_\text{PBH}^\text{peak}$&$f_c/\text{Hz}$\\
		\hline
    	H1  &$1.14\times 10^{-2}$ & $1.56\times 10^{-13}$ &$2.05\times10^{-2}$ &$8.11\times 10^{-3}$\\

    	H2 &$1.11\times 10^{-2}$ &$2.48\times 10^{-13}$ &$7.66\times10^{-3}$ &$6.40\times 10^{-3}$\\
    	
    	Hr &$1.27\times 10^{-2}$ &$5.72\times 10^{-13}$ &$4.52\times10^{-1}$ &$4.30\times 10^{-3}$\\\hline
		
    	T1 &$1.21\times 10^{-2}$&$7.85\times 10^{-13}$&$9.15\times10^{-2}$&$3.53\times 10^{-3}$\\
      	
    	T2 &$1.10\times 10^{-2}$&$1.34\times 10^{-13}$&$7.12\times10^{-3}$&$9.13\times 10^{-3}$\\
		\hline
		\hline
	\end{tabular}
	\caption{The results for the primordial power spectrum,
the peak mass and abundance of PBH
and the peak frequency of SIGWs with the chosen parameter sets shown in Table \ref{t1}. H represents the Higgs potential and T represents the T-model, 1 represents the case $n=1/3$ and 2 represents the case $n=2/3$. H1 means the model with Higgs potential and the power law potential $U(\Phi)=U_0\Phi^n$ with $n=1/3$. Hr represents the Higgs model with the running of the coupling constant $\lambda$.}
\label{t2}
\end{table}

\begin{figure}[htp]
	\centering
	\includegraphics[width=0.95\columnwidth]{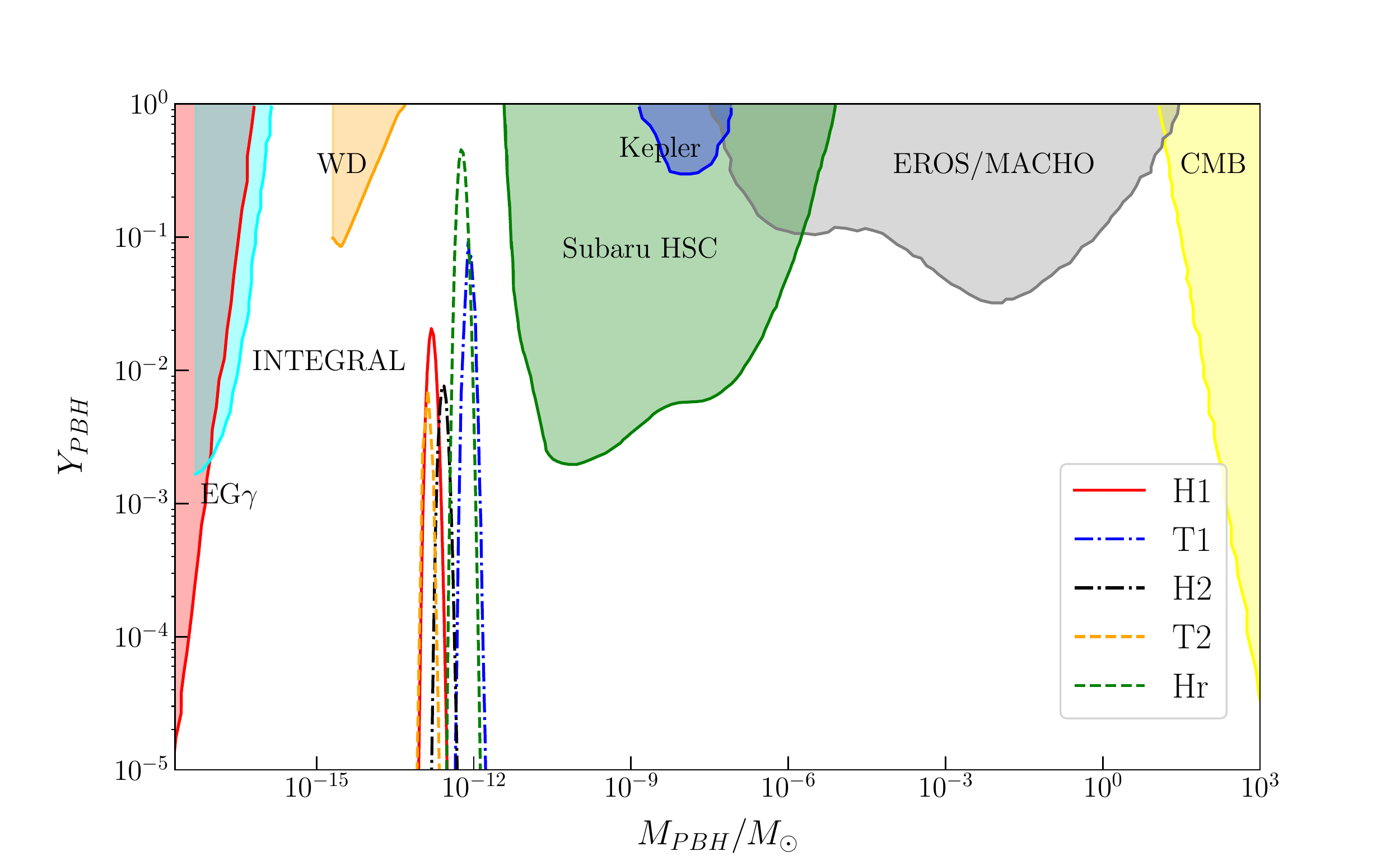}
	\caption{The results for the PBH abundance. The shaded
    regions show the observational constraints on the
    PBH abundance.
   The yellow region from accretion constraints by CMB \cite{Ali-Haimoud:2016mbv,Poulin:2017bwe},
    the red region from extragalactic gamma-rays by PBH evaporation (EG$\gamma$) \cite{Carr:2009jm}, the cyan region from galactic center 511 keV gamma-ray line (INTEGRAL) \cite{Laha:2019ssq,Dasgupta:2019cae,Laha:2020ivk}, the orange region from white dwarf explosion (WD) \cite{Graham:2015apa},
    the green region from microlensing events with Subaru HSC \cite{Niikura:2017zjd},
    the blue region from the Kepler satellite \cite{Griest:2013esa},
    and the gray region from the EROS/MACHO \cite{Tisserand:2006zx}.
    The models are the same as those in Fig. \ref{p1}.}
	\label{p2}
\end{figure}

When the overdense region generated by the primordial curvature perturbations reenters the horizon in radiation era, it can be a seed to cause  gravitational collapse to form PBHs.
The current fractional energy density of PBHs with mass $M$ to DM is \cite{Carr:2016drx,Gong:2017qlj}
\begin{equation}
\label{fpbheq1}
\begin{split}
Y_{\text{PBH}}(M)=&\frac{\beta(M)}{3.94\times10^{-9}}\left(\frac{\gamma}{0.2}\right)^{1/2}
\left(\frac{g_*}{10.75}\right)^{-1/4}\\
&\times \left(\frac{0.12}{\Omega_{\text{DM}}h^2}\right)
\left(\frac{M}{M_\odot}\right)^{-1/2},
\end{split}
\end{equation}
where $M_{\odot}$ is the solar mass, $\gamma= 0.2$ \cite{Carr:1975qj},
$g_*$ is the effective degrees of freedom at the formation time,
$\Omega_{\text{DM}}$ is the current
energy density parameter of DM,
the fractional energy density of PBHs
at the formation is related to the power spectrum of the primordial curvature perturbations as \cite{Young:2014ana, Ozsoy:2018flq,Tada:2019amh}
$$\beta(M) \approx \sqrt{\frac{2}{\pi}}\frac{\sqrt{P_{\zeta}}}{\mu_c}
\exp\left(-\frac{\mu_c^2}{2P_{\zeta}}\right),$$
where $\mu_c=9\delta_c/4$ and $\delta_c$ is the critical density perturbation for the PBH formation.
We take $\Omega_{\text{DM}}h^2=0.12$ \cite{Aghanim:2018eyx}
and $\delta_c=0.4$ \cite{Musco:2012au,Harada:2013epa,Tada:2019amh,Escriva:2019phb,Yoo:2020lmg}.
Substituting the power spectrum into Eq. \eqref{fpbheq1}, we get the abundance
of PBH DM and the result is shown in Fig. \ref{p2}.
We also show the peak mass
and the peak abundance of PBH DM in Table \ref{t2}.
It is clear that Higgs fluctuations seed the production of the PBH DM compatible with observation.
We may wonder that the large enhancement on the curvature power spectrum leads to large non-Gaussianities that affect the production of PBH DM.
It was shown in \cite{Zhang:2020uek} that the non-Gaussianities of Higgs fluctuations are small at both the CMB and peak scales
even though they reach the order one at scales larger than the peak scales,
so the effect of non-Gaussianities on the production of PBH DM is negligible.

\begin{figure}[htp]
	\centering
	\includegraphics[width=0.95\columnwidth]{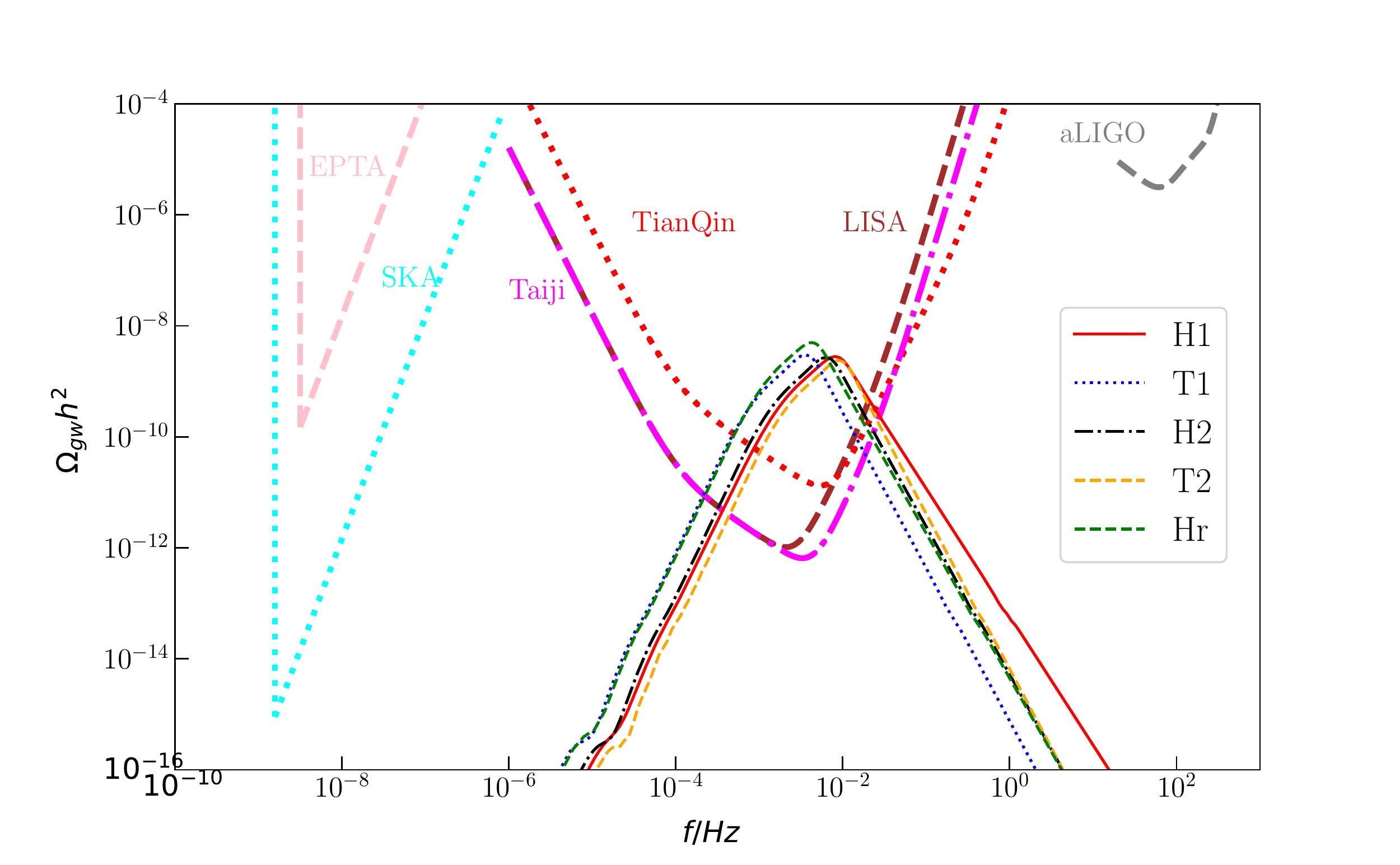}
	\caption{The results for SIGWs along with the sensitivity curves for different GW detectors.
	The pink dashed curve denotes the EPTA limit \cite{Ferdman:2010xq,Hobbs:2009yy,McLaughlin:2013ira,Hobbs:2013aka} ,
    the cyan dotted curve denotes the SKA limit \cite{Moore:2014lga},
    the red dot-dashed curve in the middle denotes the TianQin limit \cite{Luo:2015ght},
    the dotted magenta curve shows the Taiji limit \cite{Hu:2017mde},
    the brown dashed curve shows the LISA limit \cite{Audley:2017drz},
    and the gray dot-dashed curve denotes the aLIGO limit \cite{Harry:2010zz,TheLIGOScientific:2014jea}.
 The models are the same as those in Fig. \ref{p1}.}
	\label{p3}
\end{figure}

Accompanied by the production of PBHs,
the scalar perturbations can induce SIGWs during radiation.
The equation for the Fourier components of the second order
tensor perturbations $h_{\bm{k}}$ is
\cite{Ananda:2006af,Baumann:2007zm}
\begin{equation}
\label{eq:hk}
h''_{\bm{k}}+2\mathcal{H}h'_{\bm{k}}+k^2h_{\bm{k}}=4S_{\bm{k}},
\end{equation}
where $h'_{\bm{k}}=dh_{\bm{k}}/d\eta$, the scalar source
\begin{equation}
\label{hksource}
\begin{split}
S_{\bm{k}}=\int \frac{d^3\tilde{k}}{(2\pi)^{3/2}}e_{ij}(\bm{k})\tilde{k}^i\tilde{k}^j
\left[2\Phi_{\tilde{\bm{k}}}\Phi_{\bm{k}-\tilde{\bm{k}}}
+\frac{1}{\mathcal{H}^2} \right.\\
\left. \times\left(\Phi'_{\tilde{\bm{k}}}+\mathcal{H}\Phi_{\tilde{\bm{k}}}\right)
\left(\Phi'_{\bm{k}-\tilde{\bm{k}}}+\mathcal{H}\Phi_{\bm{k}-\tilde{\bm{k}}}\right)\right],
\end{split}
\end{equation}
$\mathcal{H}= 1/\eta$, $e_{ij}(\bm{k})$ is the polarization tensor, the Bardeen potential $\Phi_{\bm{k}}=\Psi(k\eta)\phi_{\bm{k}} $,
the transfer function $\Psi$ in the radiation era is
\begin{equation}
\label{transfer}
\Psi(x)=\frac{9}{x^2}\left(\frac{\sin(x/\sqrt{3})}{x/\sqrt{3}}-\cos(x/\sqrt{3})\right),
\end{equation}
and $\phi_{\bm{k}}$ is related with $P_\zeta$ as
\begin{equation}
\label{phikeq4}
\langle\phi_{\bm{k}}\phi_{\tilde{\bm{k}}}\rangle
=\delta^{(3)}(\bm{k}+\tilde{\bm{k}})\frac{2\pi^2}{k^3}\left(\frac{2}{3}\right)^2 P_\zeta(k).
\end{equation}
The power spectrum of the SIGWs is defined as
\begin{equation}
\label{eq:pwrh}
\langle h_{\bm{k}}(\eta)h_{\tilde{\bm{k}}}(\eta)\rangle
=\frac{2\pi^2}{k^3}\delta^{(3)}(\bm{k}+\tilde{\bm{k}})P_h(k,\eta),
\end{equation}
and the fractional energy density is
\begin{equation}
\label{density}
\Omega_{\mathrm{GW}}(k,\eta)=\frac{1}{24}\left(\frac{k}{aH}\right)^2\overline{P_h(k,\eta)}.
\end{equation}
Combining Eqs. \eqref{eq:hk}-\eqref{density} and the primordial power spectrum $P_\zeta$, we obtain $\Omega_{\mathrm{GW}}$ and the result is shown in Fig. \ref{p3}, which tells us that the produced SIGWs can be detected in the future space-based GW observatories like LISA, Taiji and TianQin. Since the Non-Gaussianitiy is small at the peak scale, so its effect on SIGWs is negligible \cite{Zhang:2020uek}.

Now we show that a significant abundance of PBH DM  and observable SIGWs can be produced even the running of the coupling constant $\lambda$ is considered.
The model is labeled as Hr.
The running self-coupling has a minimum at the energy scale $\phi=\mu\sim10^{17-18}~\text{Gev}$,
around which it can be expanded as  $\lambda(\phi)=\lambda_0+b_0 \ln^2(\phi/\mu)$ with $b_0\approx2.3\times10^{-5}$ \cite{Isidori:2007vm,Hamada:2013mya,Hamada:2014iga,Bezrukov:2014bra,Ezquiaga:2017fvi}.
Taking into consideration the energy scale $\mu$,
we rewrite the  function $f(\phi)$ as $f(\phi)=\tilde{f}_0(\phi/\mu)^{22}$.
If we choose
\begin{equation}
\lambda_0=6.20\times10^{-6},\quad \mu=0.1, \quad \tilde{f}_0=10,
\end{equation}
and other parameters as shown in Table \ref{t1},
we get $n_s=0.9676$,  $r= 0.054$, and $N=60.2$ which are consistent with the observational data.
The results for the power spectrum, the PBH abundance and SIGWs
are shown in Figs. \ref{p1}, \ref{p2}, and \ref{p3} respectively.

To show that the mechanism can give different $n_s$ and $r$, we take the Higg
potential with the power law $U(\Phi)=U_0\Phi^{2/3}$  as an example and label it as H2.
Taking the parameter set  in Table \ref{t1},
we get $n_s=0.9697$, $r=0.0641$ and $N=64.2$.
The results for the power spectrum, the PBH abundance and SIGWs
are shown in Figs. \ref{p1}, \ref{p2}, and \ref{p3} respectively.

In order to show that our treatment is not specific to a particular potential, we generalize our discussion to the T-model.
For the T-model \cite{Kallosh:2013maa,Kallosh:2013hoa,Yi:2016jqr}
\begin{equation}\label{tmodelp}
V=V_0\tanh^{2m}\left(\frac{\phi}{\sqrt{6 \alpha}}\right),
\end{equation}
we can derive the attractor $n_s=1-2/N$ and $r=12/N^2$ which are consistent
with Planck 2018 data for $N=50-60$.
In our mechanism, we choose the function $f(\phi)$ as
\begin{equation}\label{tmodel:fn}
  f(\phi)=f_0~ \text{sech}^4\left(\frac{\phi}{\sqrt{6\alpha}}\right) \tanh^{-2+4m/n}\left(\frac{\phi}{\sqrt{6\alpha}}\right),
\end{equation}
where
\begin{equation}\label{tmodel:fn0}
f_0=\frac{2}{3\alpha} \left(\frac{V_0}{U_0}\right)^{2/n}\frac{m^2}{n^2}.
\end{equation}
In this paper we take $f_0= 36$.
The T-model with $m=1/6$ and $ \alpha=1$ combined with the power law $U(\Phi)=U_0\Phi^{1/3}$ is labelled as T1
and the T-model with $m=1/3$ and $ \alpha=1$ combined with the power law $U(\Phi)=U_0\Phi^{2/3}$ is labelled as T2.
The model parameters and the results are shown in Tables \ref{t1} and \ref{t2}
and Figs. \ref{p1}, \ref{p2}, and \ref{p3}.

From these results, we see that
our mechanism is general and appropriate to  both the Higgs field and the T-model.
We have shown that both models are consistent with Planck 2018 data.
A significant abundance of PBH DM and SIGWs can be produced in our mechanism
which is expected to be detectable by LISA/TianQin/Taiji detectors,
and the mechanism is not restricted  to the two models discussed.

\section{conclusion}

In this work we proposed a novel mechanism to resolve the contradiction in the original k/G inflation of simultaneously requiring  the enhancement of the curvature perturbations at small scales and keeping the model predictions in consistent with CMB observations
at large scales for Higgs field.  We found that with this mechanism Higgs field  inflationary model becomes viable.
In our method, the function $G_a(\phi)$ peaks near
the end of inflation thereby the enhancement of the power spectrum
happens at small scales only. Such peak contributes about
20 $e$-folds during the enhancement.
To keep the number of $e$-folds to be $50-60$, the field
value $\phi_*$ at the horizon exit moves closer to the field
value at the end of inflation and the slow-roll contributions to $n_s$ and $r$ are changed.
Away from the peak, the function $G_a(\phi)$ is negligible
and the usual slow-roll inflation applies,
the noncanonical term with the function $f(\phi)$ ensures
the power spectrum at large scales to be
consistent with CMB observations.
In our mechanism,
the observables $n_s$ and $r$ are not sensitive to the inflaton potential, where both Higgs potential and the general T-models can be employed to describe the Planck 2018 observations.
The mechanism does not restrict the functions $G_a(\phi)$
and $f(\phi)$ in the noncanonical kinetic term  and the inflationary potential to the particular forms used in this paper,
other forms are permitted.
Different peak shape, different peak mass for PBH DM
and different peak frequency for SIGWs are all possible.

The Higgs boson of the standard model of particle physics
is responsible not only for the masses of elementary particles,
but can act as an inflaton to drive inflation to meet CMB measurements.
Furthermore, we have shown that it can explain the DM content of our universe in the form of PBHs.
The SIGWs induced by the large first order Higgs fluctuations
at small scales can be observed by the space-based GW observatories, such as LISA, Taiji and TianQin.
Future astronomical observations can grasp more signatures of Higgs field through PBHs DM and SIGWs.

\begin{acknowledgments}
This research is supported in part by the National Natural Science
Foundation of China under Grant Nos. 11875136 and 12075202, and
the Major Program of the National Natural Science Foundation of China under Grant No. 11690021.
Z. Yi acknowledges support by China Postdoctoral Science Foundation Funded Project under Grant No. 2019M660514.
Z. Zhu is supported by the National Natural Science Foundation of China under Grants Nos. 11633001, 11920101003 and 12021003, the Strategic Priority Research Program of the Chinese Academy of Sciences, Grant No. XDB23000000 and the Interdiscipline Research Funds of Beijing Normal University.
\end{acknowledgments}


%

\end{document}